\documentclass[jmp,amsmath,amssymb,preprint]{revtex4-1}
\usepackage{graphicx}
\usepackage{dcolumn}
\usepackage{bm}

\begin{document}

\title{Magnon-drag thermopile}

\author {Marius V. Costache$^1$}
\author {German Bridoux$^1$}
\author {Ingmar Neumann$^1$}
\author {Sergio O. Valenzuela$^{1,2,3}$}
\affiliation{$^1$Catalan Institute of Nanotechnology (ICN-CIN2), Barcelona E-08193, Spain}
\affiliation{$^2$Universitat Aut\`{o}noma de Barcelona (UAB), Barcelona E-08193, Spain}
\affiliation{$^3$Instituci\'{o} Catalana de Recerca i Estudis Avan\c{c}ats (ICREA), Barcelona E-08010, Spain}

\date{\today}


\begin{abstract}
\noindent \textbf{Thermoelectric effects in spintronics \cite{johnson1987} are gathering increasing attention as a means of managing heat in nanoscale structures and of controlling spin information by using heat flow \cite{mccann2002,gravier2006,uchida2008,uchida2010,jaworski2010,slachter2010,ansermet2010,jansen2011,walter2011}.
Thermal magnons (spin-wave quanta) are expected to play a major role \cite{mccann2002,uchida2010,xiao2010,avery2011}, however, little is known about the underlying physical mechanisms involved. The reason is the lack of information about magnon interactions and of reliable methods to obtain it, in particular for electrical conductors because of the intricate influence of electrons \cite{xiao2010,sinova2010}.
Here, we demonstrate a conceptually new device that allows us to gather information on magnon-electron scattering and magnon-drag effects. The device resembles a thermopile \cite{disalvo1999} formed by a large number of pairs of ferromagnetic wires placed between a hot and a cold source and connected thermally in parallel and electrically in series. By controlling the relative orientation of the magnetization in pairs of wires, the magnon-drag can be studied independently of the electron and phonon-drag thermoelectric effects. Measurements as a function of temperature reveal the effect on magnon drag following a variation of magnon and phonon populations. This information is crucial to understand the physics of electron-magnon interactions, magnon dynamics and thermal spin transport.
}
\end{abstract}

\maketitle

A ferromagnet subject to a temperature gradient contains a higher density of magnons in the hotter region, which by diffusive motion move towards the cooler region. Because of electron-magnon collisions, the diffusion of magnons may add a magnon-drag contribution to the Seebeck coefficient of the ferromagnet. Owing to the analogy between electron scattering processes by phonons and magnons, it has long been accepted that the theory of magnon drag should follow that of phonon drag \cite{bailyn1962}, but this has hardly been observed \cite{grannemann1976,cox1967,avery2011}. Indeed, it is remarkable the scarce evidence of magnon-drag effects in 3\textit{d} ferromagnets, even after more than 40 years of intense research on these materials. The presence of magnon drag is usually inferred indirectly from measurements of the temperature variation of the Seebeck effect, whose interpretation is difficult and masked by phonon and electron contributions. In principle, measurements as a function of magnetic field should provide information on magnon drag but the experiments have been performed above a few Tesla at low ($<$ 4 K) temperatures \cite{grannemann1976}. This is due to the low sensitivity of such measurements and in order to avoid the presence of magnetic domains and intrinsic anisotropic magnetoresistance effects that are common at lower magnetic fields \cite{grannemann1976,racquet2002,mihai2008}. Information at low magnetic fields has therefore proven even more difficult to obtain, even though it is particularly relevant for ongoing thermoelectric experiments in magnetic systems \cite{mccann2002,uchida2008,uchida2010,jaworski2010,slachter2010,ansermet2010,jansen2011,walter2011,avery2011,hinzke2011}. Overall, the experimental results have been interpreted only qualitatively and comparison with theoretical modeling has not been possible.

Our measurement scheme (Fig. 1) allows us to study magnon-drag effects at low magnetic fields ($\sim$ 0.1 T) isolated from such spurious and competing phenomena. The scheme relies on the fact that a magnetic field parallel (antiparallel) to the magnetization leads to a decrease (increase) of the magnon population, as demonstrated by magnetoresistance measurements \cite{racquet2002,mihai2008} (Fig. 1a). This can be understood by inspecting the magnon dispersion relation for magnon wave vector $q$, which has the quadratic form $E(q)\approx Dq^2+g \mu_B B_{\mathrm{int}}$ \cite{racquet2002,mihai2008,gurevich1996}, where $D \sim D_0(1-d_1T^2)$ is the magnon mass renormalization, $D_0$ the zero temperature magnon mass and the $d_1T^2$ term accounts for the temperature dependence of the Fermi distribution that leads to an increase of the effective magnon mass. In the second term, $B_{\mathrm{int}}=B+\mu_0M$ corresponds to the magnetic field induction $B$ plus the ferromagnet magnetization $\mu_0M$. A change in $B$ modifies $B_{\mathrm{int}}$ and the magnon modes that are attainable at a given temperature. Therefore, a $B$ antiparallel (parallel) to $M$ should make the highest wave vector modes (un)reachable.

\begin{figure}[t]
\begin{center}\includegraphics[width=6.2in]{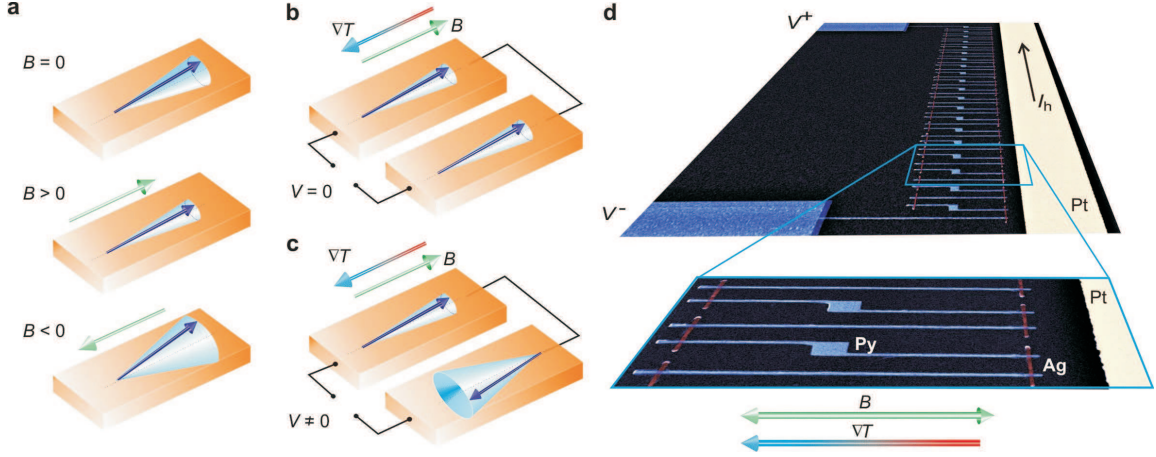}\end{center}\vspace{-5mm} \caption{\footnotesize{Magnon-drag detection principle and geometry of the device. \textbf{a}, The magnon population in ferromagnetic wires is represented schematically by an arrow with a given precession angle. A magnetic field, $B>0$, applied parallel to the magnetization, leads to a reduction/damping of the magnon population (represented by a reduction in the precession angle). In contrast, a magnetic field $B<0$, antiparallel to the magnetization, results in an increase of the magnon population (increase in the precession angle). \textbf{b, c}, Magnon-drag measurements. The wires are connected thermally in parallel and electrically in series. For parallel magnetization orientation (\textbf{b}), the thermoelectric voltage is zero because the contribution from each wire is the same and has opposite sign. For the antiparallel configuration (\textbf{c}), in the presence of $B$, a nonzero thermoelectric voltage is measured due to the difference in magnon-drag effect, directly related to the difference in magnon populations induced by $B$. \textbf{d}, Scanning electron microscope image (SEM) of a typical device. A large number of pairs of NiFe wires (blue) are connected in series with Ag wires (red). A wide Pt wire (yellow) serves as a heater to generate a thermal gradient, $\nabla T$.}
} \label{fig1}
\end{figure}

In our devices (Fig. 1d), we use narrow NiFe wires (20 nm thick, 30 nm wide and 5 $\mu$m long) that contain a single magnetic domain. A large number $N$ of pairs of wires are placed between a hot and a cold source and are connected thermally in parallel and electrically in series in the presence of a magnetic field along the magnetization direction. Each wire in a pair is identical to the other except for a short broader region in one of them. This results in different coercive fields ($B_{c1} < B_{c2}$) that allow us to control the relative orientation of the wire magnetizations.

The device is effectively a magnon-drag thermopile. When the magnetizations are in the parallel configuration, the measured thermoelectric voltage $V$ is zero and independent of magnetic field because the contributions of each wire are of the same magnitude and have opposite sign (see Fig. 1b). However, when the magnetizations are in the antiparallel configuration the measured thermoelectric voltage is determined by the difference in magnon populations induced by the magnetic field (Fig. 1c). Other contributions to the thermoelectric voltage, related to electrons and phonons, are independent of the magnetization orientation and cancel out in each pair. The effect of the Lorentz force, relevant at high fields, is also negligible below one Tesla \cite{racquet2002}. As in a conventional thermopile, the thermopower for a single pair $S$ is multiplied by the number of pairs, which in our case is $N=20$. The device thus generates the magnon-drag thermopower, $S^N =N S$, which would be undetectable using a single pair. Note that $S$ is an intrinsic property of the ferromagnet in contrast to conventional thermopower measurements, which include the thermopower of both the material of interest and the material of the measurement electrodes.

We prepare the devices using two electron-beam lithography steps and a two-angle shadow-mask evaporation technique \cite{SOV2007,SOV2009,MVC2010}. First a Pt heater is defined. Then, the thermopile is created by depositing sequentially, using a shadow mask, the NiFe wires and transverse Ag wires that connect the NiFe wires electrically in series (Supplementary Figure 1).

We first characterize our devices electrically. We use magnetoresistance (MR) measurements as in Ref. \cite{mihai2008} to demonstrate the change in magnon population as a function of magnetic field and to obtain the temperature variation of the magnon mass. Fig. 2a shows typical measurements at room temperature. The magnetic field is swept along the axis of the ferromagnets. At large enough negative $B$, the magnetizations of the wires are in a parallel configuration. As $B$ is swept from negative to positive, the magnetizations of the wires with lower coercive field ($B_{c1}$) reverse and the device switches to an antiparallel configuration (see inset Fig. 2a). As $B$ is further increased, the magnetizations of the remaining wires also reverse (at $B_{c2}$) and a parallel configuration is recovered. An analogous sequence occurs when sweeping the magnetic field from positive to negative starting at large positive $B$.

In the parallel configuration, the device presents a linear non-saturating negative MR, consistent with recent reports on thin films \cite{racquet2002,mihai2008}. This decrease is attributed to the reduction of electron-magnon scattering processes due to the decrease in the magnon population. In the antiparallel configuration, in contrast, the resistance shows a plateau with $B$. This is so because for the wire with $M$ antiparallel to $B$, MR increases with $B$, whereas for the wire with $M$ parallel to $B$, MR decreases with $B$, therefore, their contributions to the total MR tend to cancel each other out (Fig. 2a inset).

The magnetoresistance is strongly temperature dependent. Fig. 2b shows the resistance difference $\Delta R$ between parallel and antiparallel magnetization configurations at fixed $B =0.125$ T. At low enough temperatures, the thermally excited population of magnons drops significantly resulting in a drop in $\Delta R$. From these measurements, both $D_0$ and $d_1$ are determined \cite{mihai2008} (Fig. 2b). By introducing a value of $M = 1$ T \cite{costache2006}, we obtain a very good fit with $d_1 \approx 4 \times 10^{-6}$ K$^{-2}$, in agreement with previous studies in Ni and Fe films \cite{racquet2002}.

\begin{figure}
\begin{center}\includegraphics[width=3.2in]{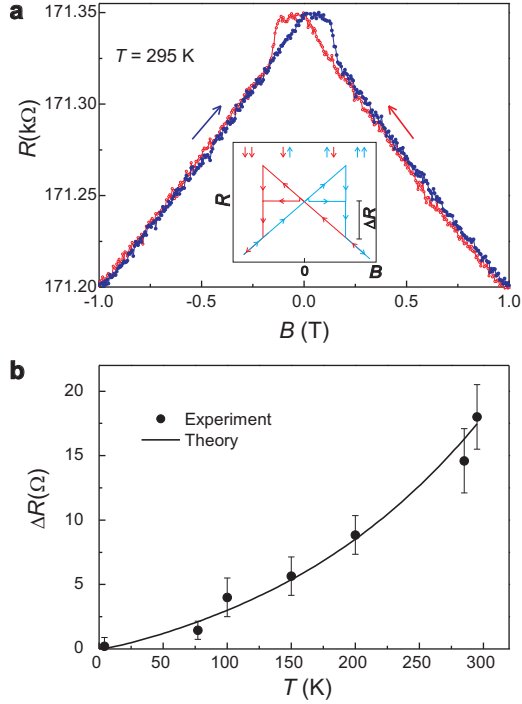}\end{center}\vspace{-15mm} \caption{\footnotesize{Magnon-induced magnetoresistance (MR). \textbf{a}, Typical magnetoresistance measurements at 295 K. The blue and red arrows indicate the $B$ sweep direction. A linear decrease of the magnetoresistance is attributed to magnon damping. The hysteresis around zero field occurs when the device switches from parallel to antiparallel configurations. Inset, schematic illustration of the MR hysteresis at low $B$. The $B$ swept direction is marked by arrows. The top arrows indicate the relative magnetization orientation of the wires. A plateau is observed for $B_{c1}<B<B_{c2}$ in the antiparallel configuration. \textbf{b}, Change in $R$, $\Delta R$, between parallel and antiparallel configurations at $B = 0.125$ T as a function of temperature. The solid line represents a fit to a model based on magnon-electron interactions \cite{racquet2002,mihai2008}. The error bars indicate noise estimates from the raw measurements (\textbf{a}) ($\pm$ 2 s.d.).}
} \label{fig2}
\end{figure}

Having demonstrated that the MR induced by magnons presents the expected magnetic field and temperature responses, we perform thermopower measurements, $S^N = V$/$\Delta T$, in the same device in order to study magnon-drag effects. For this purpose, we measure the dc voltage, $V$, between the $V^{+}$ and $V^{-}$ electrodes that results from a temperature difference $\Delta T$ between the two ends of the wires (Fig. 1d). The temperature difference is determined in an identical sample with built-up Pt thermometers at the locations of the Ag wires \cite{small2003,boukai2008}. Figure 3a shows typical data at 50 K and 105 K. The magnetic field is again swept along the wires. The thermoelectric signal, $S^N$, only appears in the antiparallel configuration, it is linear with $B$, and extrapolates to zero at $B=0$. At $B=0$, the magnetization configuration is always parallel in these measurements. However, Fig. 3b shows that both parallel and antiparallel configurations are possible at $B=0$ and that they can be prepared in a controlled way. The antiparallel configuration is achieved by reversing the sweep direction of $B$ at $B_{c1}<B<B_{c2}$. In this situation, the voltage follows the linear dependence until $B=-B_{c1}$, where the parallel configuration is recovered.

\begin{figure}
\includegraphics[width=4in]{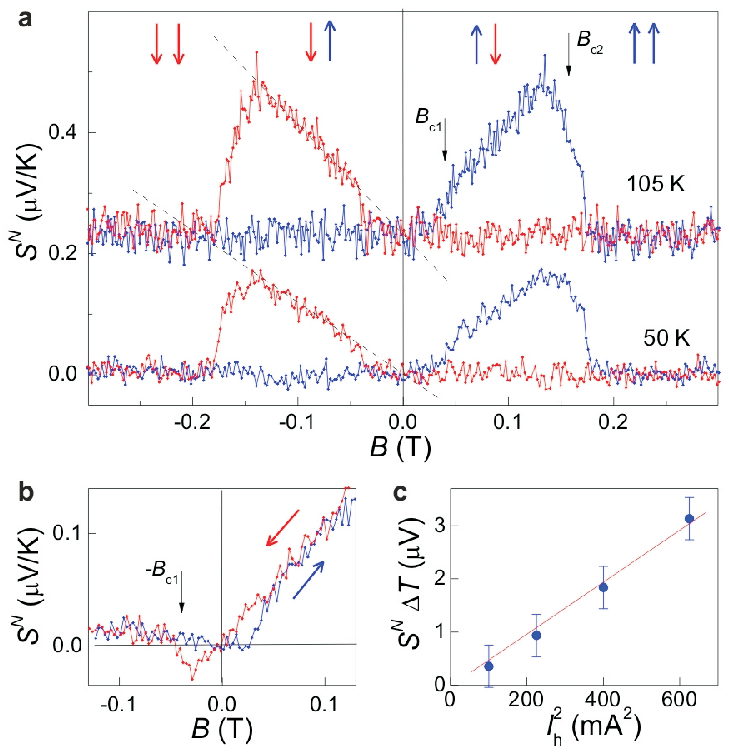}\vspace{-10mm} \caption{\footnotesize{ Magnon-drag. \textbf{a}, Thermopower, $S^N$, as a function of $B$ at 50 K and 105 K. The red (blue) lines correspond to the negative (positive) sweep direction of $B$. A dc voltage $V$ is observed in the antiparallel configuration that varies with $B$. $V$ varies linearly with $|B|$ and extrapolates to zero for $B=0$. For clarity, the 105 K data is offset upwards and a constant background was subtracted. \textbf{b}, Minor hysteresis loop that preserves the antiparallel configuration at $B=0$. The $B$ sweep direction is marked by arrows. Starting at $B < 0$, the sweep direction is reversed at +0.12 T $<B_{c2}$. The magnetic configuration remains antiparallel until $B=-B_{c1}$ and a linear behaviour is observed for $-B_{c1}< B <B_{c2}$. $T$ = 50 K. \textbf{c}, Thermoelectric voltage, $V=S^N \Delta T$ as a function of the heater current, $I_h$. $T$ = 50 K. The error bars indicate noise estimates from the raw measurements of $V$ ($\pm$ 2 s.d.).}
} \label{fig3}
\end{figure}

The minor hysteresis loop in Fig. 3b helps verify that $S^N$ indeed extrapolates to zero at $B=0$. Note that the voltage at $B=0$ is zero independently of the magnetization configuration, which is evidence that $S^N$ is due to a difference in magnon population induced by the applied field. The thermoelectric origin of the signal is further proved in Fig. 3c, where $V = S^N \Delta T$ is found to be proportional to the square of the heater current, thus to the temperature gradient.

Measurements of the magnon drag as a function of temperature should reflect the change in the electron-magnon and magnon-magnon collisions resulting from the variation of the magnon population. The thermopower at $B=$ 0.125 T as a function of temperature is shown in Fig. 4a. It first increases with temperature and then, around $T_{\textrm{peak}} = 180$ K (about one fifth of the NiFe Curie temperature), it starts decreasing rapidly. This behaviour agrees with the expectations from the phonon-drag analogy. Following this analogy, the initial increase is the result of an increase of the magnon population, whereas the subsequent decrease is due to a larger probability of magnon-magnon or magnon-phonon collisions that reduce the momentum transfer from magnons to electrons.

Quantitatively, the magnon-drag contribution to the thermoelectric voltage depends on the extent of magnon momentum transfer to the electrons as compared to the transfer to other magnons, phonons and impurities. Following the theory of phonon drag \cite{Ziman}, we assume that the magnon-drag magnitude is proportional to the probability of magnon-electron interaction, $P_{m,e}$, divided by the probability of a magnon interacting with any particle including electrons,\textit{ i.e.} $P_{m,x}$+$P_{m,e}$. Written in terms of interaction times, $\tau_{m,e}$, $\tau_{m,x}$, the magnon-drag contribution to the thermopower is $S^{MD}\propto P_{m,e}/(P_{m,x}+P_{m,e})\propto \tau_{m,x}/(\tau_{m,x}+\tau_{m,e})$. By considering the drift of electron and magnons and the quadratic dispersion relation for magnons, $S^{MD}$ (for a single wire) can be calculated \cite{grannemann1976,gurevich1996},

\begin{equation}
S^{MD}(B,T)=\frac{1}{n'_{e}e}\frac{k^{5/2}_{B} T^{3/2}F(y)}{6\pi^{2} D^{3/2}} \left(\frac
{\tau_{m,x}}{\tau_{m,x}+\tau_{m,e}}\right),
\label{magnon-drag}
\end{equation}
\\
\noindent where $y=(g\mu_{B}B_\mathrm{int}/k_{B}T)$ and $F(y)$ is the `quenching function', which gives the magnetic field contribution to the magnon-drag effect (see Supplementary Information), and $D$ is the exchange stiffness constant obtained from the MR analysis. For $y < 0.1$, \textit{i.e.} the low field regime ($B \lesssim 3$ T at 50 K), $F(y)$ is linear, which explains the linear dependence in Figs. 3a and b. In equation (1), $n'_{e}= n_e/\alpha$, where $n_e$ is the number of conduction electrons per unit volume and $\alpha$ is a phenomenological parameter of order 1. This parameter accounts for the fraction of electrons that are involved in collisions with magnons. It reflects the fact that electrons with high mobility and strong 4$s$ character will likely dominate the collisions when compared to the low mobility electrons with 3$d$ character \cite{grannemann1976}.

\begin{figure}
\includegraphics[width=3.2in]{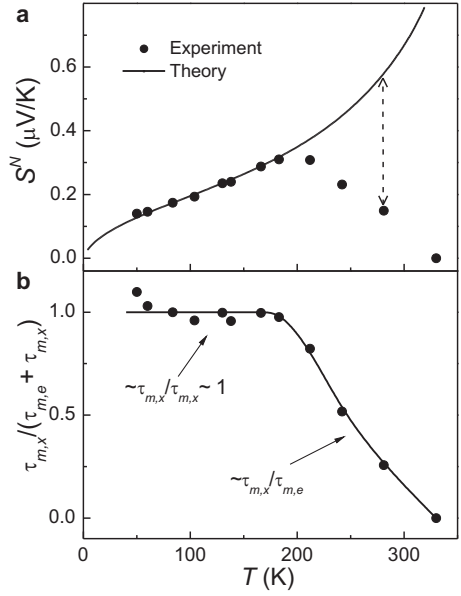}\vspace{-10mm} \caption{\footnotesize{Temperature variation of the magnon-drag effect. \textbf{a}, $S^N$ as a function of $T$. Experimental data is shown with solid dots. The black solid line represents a best fit to equation (1) in the low temperature region. \textbf{b}, Extracted $\tau_{m,x}/\tau_{m,e}$ using equation (1) and data in \textbf{a}. $B=0.125$ T.}
} \label{fig4}
\end{figure}

At low temperatures, the population of magnons is low and therefore we expect that $\tau_{m,e} \ll \tau_{m,x}$, that is $\tau_{m,x}/(\tau_{m,x}+\tau_{m,e})\approx 1$. The momentum transfer between magnons and electrons will occur with high probability and $\alpha$ becomes the only fitting parameter in equation (1). To fit our experimental results, we note that the above model predicts that the measured thermoelectric response in the antiparallel configuration at temperature $T$ and magnetic field $B$ is $S^N = V/\Delta T = N [S^{MD}(T,-B)-S^{MD}(T,+B)]$. The two terms account for the wires whose magnetization is antiparallel to $B$ (first term) and parallel to $B$ (second term). The best fit to this expression shows excellent agreement with our data (line in Fig. 4), specially considering that it involves only one fitting parameter ($\alpha \approx 3$). This strongly supports the assumption that $\tau_{m,e} \ll \tau_{m,x}$ below $T_{\textrm{peak}}$ and our model of magnon drag.

The signal decrease at high temperatures is the result of magnon-magnon and magnon-phonon interactions. At the peak $\tau_{m,x}\sim \tau_{m,e}$, whereas at higher temperatures magnon collisions not involving electrons become more frequent such that $\tau_{m,x} \ll \tau_{m,e}$ and $S^{MD} \propto \tau_{m,x}/\tau_{m,e}$. Around room temperature, the magnon momentum loss turns out to be large enough to almost completely suppress the magnon drag effect. If we extrapolate the low temperature fit to temperatures above the peak, we can estimate the ratio $\tau_{m,x}/\tau_{m,e}$ that gives place to the suppression of the thermopower indicated by the dotted arrow in Fig. 4a (see Fig. 4b). This ratio provides an important insight into the thermal dynamics of magnons. However, further theoretical studies are required to model the experimental results in this temperature range, in particular, to predict the value of $\alpha$, the magnitude of the drop at large temperatures, or even the position of the peak $T_{\textrm{peak}}$.

We have thus presented magnon-drag measurements in NiFe using a thermopile-like device, and modeled the effect at low temperatures using a single fitting parameter. Our results not only demonstrate the existence of magnon-drag in NiFe but also ratify a technique that provides reliable information on magnon-drag at low magnetic fields and in a broad temperature range that was not accessible before. Although it is not possible with our current analysis to separate magnon scattering times with magnons and phonons, these results have important implications for thermoelectricity in magnetic structures \cite{uchida2008,uchida2010,jaworski2010,slachter2010,ansermet2010,jansen2011,avery2011}. For example, the difference in the effective temperatures of magnons and electrons and their interactions are gaining relevance for modeling novel thermoelectric effects. The dominant interaction of magnons below $T_{\textrm{peak}}=180$ K is with electrons and not phonons (or other magnons) and therefore the temperature difference would likely vanish below $T_{\textrm{peak}}$.

\vspace{5mm}

\noindent \textbf{Acknowledgments} We gratefully acknowledge discussions with I. M. Miron and J. Van de Vondel. We thank S. Alvarado, A. Bachtold, O. Fesenko and P. Gambardella for a critical reading of the manuscript. This research was supported by the Spanish Ministerio de Ciencia e Innovaci\'on, MICINN (MAT2010-18065) and by the European Community's Seventh Framework Programme (FP7/2007-2013) under grant agreement NANOFUNCTION n$^\circ$257375.

\end{document}